\documentstyle[epsfig,12pt]{article}

\newcommand{\bce}{\begin{center}} 
\newcommand{\ece}{\end{center}}
\newcommand{\beq}{\begin{equation}}
\newcommand{\eeq}{\end{equation}}
\newcommand{\bea}{\vspace{0.25cm}\begin{eqnarray}}
\newcommand{\eea}{\end{eqnarray}}

\newcommand{\bk}{{\bf k}}

\newcommand{\ba}{\begin{array}}
\newcommand{\ea}{\end{array}}

\newcommand{\ket}[1]{| {#1} \rangle}

\newcommand{\bkappa}{\mbox{\boldmath ${\kappa}$}}

\def\lsim{\mathrel{\rlap{\lower4pt\hbox{\hskip1pt$\sim$}}
    \raise1pt\hbox{$<$}}}         
\def\gsim{\mathrel{\rlap{\lower4pt\hbox{\hskip1pt$\sim$}}
    \raise1pt\hbox{$>$}}}         

\def\beq{\begin{equation}}
\def\endeq{\end{equation}}
\def\arr{\begin{eqnarray}}
\def\endarr{\end{eqnarray}}

\makeindex

  \advance\oddsidemargin  by -1in
  \advance\evensidemargin by -1in
  \advance\topmargin      by -0.5in
\begin{document}

\vspace{2.0cm}

\begin{flushright}
\end{flushright}

\vspace{1.0cm}

\begin{center}
{\Large \bf UHE neutrinos: higher twists,
scales,  saturation}\footnote{Talk given at Diffraction 2010, 
Otranto (Lecce), Italy, September 10-15, 2010.}

\vspace{1.0cm}

{\large\bf R.~Fiore$^{1}$ and V.R.~Zoller$^{2}$}

\vspace{1.0cm}

$^1${\it Dipartimento di Fisica,
Universit\`a     della Calabria\\
and\\
 Istituto Nazionale
di Fisica Nucleare, Gruppo collegato di Cosenza,\\
I-87036 Rende, Cosenza, Italy}\\
$^2${\it
ITEP, Moscow 117218, Russia\\}
\vspace{1.0cm}

\end{center}
{\bf Abstract.} It is shown  that in the ultra-high energy neutrino interactions  
the higher twist corrections
brought about by  the non-conservation  of the  top-bottom current 
dramatically change the longitudinal structure function, $F_L$. 
To the Double Leading Log Approximation simple and numerically accurate
formulas for $F_L$ and $\sigma^{\nu N}$ are derived.

\bigskip

\section{What is UHE ?}
Neutrinos coming from  active galactic nuclei, gamma ray
 bursts \cite{Becker} and emerging in more speculative scenarios like
breakdown of Lorentz invariance and   
decays of super-massive particles have rather hard spectrum extending beyond 
$E_{\nu}\sim 10^{11}$ GeV \cite{Drees}.
These  Ultra-High Energy (UHE) neutrinos 
probe the gluon density in the target nucleon at very small values of 
Bjorken $x$ thus providing an opportunity of doing small-$x$ physics in
a new kinematical domain. The properties of 
the  neutrino-nucleon total cross section
$\sigma^{\nu N}(E_{\nu})$ at
 $E_{\nu}$ above  $10^{8}$ GeV were analyzed by many \cite{Gluck}. 

\bigskip

\section{Scales - 
prodotti tipici}

The overall hardness scale of the process $\nu N\to \mu X$ is 
usually estimated as
\bea 
Q^2\sim m_W^2. 
\label{eq:Q2MW2}
\eea
Indeed,
to the Double Leading Log Approximation (DLLA)
\bea
\sigma^{\nu N}\propto \int dQ^2 
\left({m_W^2\over m_W^2+Q^2} \right)^2
\exp\sqrt{C\log(1/x)\log\log Q^2}
\eea
and the origin of  Eq.(\ref{eq:Q2MW2})
becomes evident.
This observation entails (is based on) the smallness of the 
characteristic value of 
Bjorken  $x$ which is  $x\sim m_W^2/2m_NE_{\nu}$. 

\bigskip
 
\section{When top enters the game}
The above  estimate, 
$
Q^2\sim m_W^2,
$
is not unreasonable  only for 
light flavor currents.
The top-bottom current  needs special care.  
The phenomenon of 
Charged Current Non-Conservation 
(CCNC) pushes the hardness scale up to 
$\sim m_t^2$ \cite{FZPL}.

\bigskip

\section{$F_L$ as a carrier of CCNC effects}
Weak currents are not conserved. But in what way?
For longitudinal/scalar  W-boson the transition vertex $W_L\to t\bar b$
is $\propto \varepsilon_L^{\mu}J_{\mu}\propto\partial_{\mu}J_{\mu}\propto m_t\pm m_b$. Therefore, the observable quantity
$F_L\propto \varepsilon_L^{\mu}T^{\mu\nu}\varepsilon_L^{\nu}$ 
called the longitudinal structure function
provides a measure of the  CCNC effect. 
Here $T^{\mu\nu}$ represents the imaginary part of the forward
 scattering Compton amplitude. The longitudinal component of the
${\nu N}$ total cross section is  proportional to $F_L$.

\bigskip

\section{$F_L$ and $\kappa$-factorization}
\begin{figure}[h]
\psfig{figure=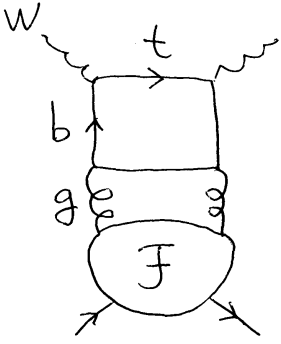,scale=0.60}
\vspace{-0.0cm}
\caption{The forward Compton scattering amplitude} 
\label{fig:fig1}
\end{figure}

The gauge invariant sum of diagrams like that shown 
in \ref{fig:fig1} results in 

\bea
{dF_{L}(x,Q^{2})\over dz d^{2}\bk }= 
{Q^2\over 4\pi^3}
 \int { d^{2}\bkappa 
\over \bkappa ^{4}}\alpha_{S}(q^{2})
{\cal F}(x,\bkappa ^{2})\left(V_S+A_S+V_P+A_P\right),
\label{eq:dsdzdk}
\eea
where ${\cal F}$ is  un-integrated gluon density,
$\bkappa$ - gluon momentum,
$z, \bk $ - Sudakov's variables of t-quark.
We find it convenient to separate contributions of  the light cone 
Fock states 
 $\ket{t\bar b}$ with angular momentum $L=0$ (S-wave)
and $L=1$ (P-wave).
The appearance of the  P-wave component  is  the manifestation of the CCNC.

\section{Higher twists. P-wave: $\ket{W}\to \ket{ t \bar b, L=1}$}
Normally, the transition of the light cone scalar $W$-boson into the P-wave 
$q\bar q^{\prime}$-state is suppressed by the factor $m_q^2/Q^2$
 \cite{FZPL}.
However, in the case at issue $m_q^2\equiv m_t^2\gg Q^2\sim m_W^2$ 
and, consequently,  there is no suppression at all.
Upon the azimuthal averaging
\begin{equation}
\langle V_P(m_t,m_b)\rangle\simeq {(m_t-m_b)^2\over Q^2}
{\bkappa^2(\bk^{4}+\varepsilon^{4}) \over  
(\bk^{2}+\varepsilon^{2})^4}
\label{eq:VP}
\end{equation}
and
$
\langle A_P(m_t,m_b)\rangle=\left(g_A/g_V\right)^2V_P(m_t,-m_b),$
where
$\varepsilon^{2}=z(1-z)Q^{2}+(1-z)m_t^{2}+zm_b^2.$

In the 
 soft gluon approximation, 
$\kappa^2\ll k^2+\varepsilon^2$, and 
 the  P-wave component 
of
\begin{equation}
F_L=F^S_L+{F^P_L} 
\label{eq:FL}
\end{equation}
is dominated by highly asymmetric configurations
with \cite{FZJL}
$$z\sim 1-{m_b^2\over m_t^2+Q^2}.$$
Therefore,
\begin{equation}
{F^P_{L}(x,Q^{2})}\simeq
{m_t^2\over {m_t^2+Q^2}}\int_{m_b^2}^{m^2_t}
{d\varepsilon^{2}\over \varepsilon^{2}} 
 {\alpha_S(\varepsilon^{2}) \over 3\pi}G(x,\varepsilon^2)
\label{eq:FLP}
\end{equation}
Note, the factor $m_t^2/(m_t^2+Q^2)$ emerges here as a
 property of the transition vertex $W\to t \bar b$ 
  rather than the property of the interaction of the
light cone $t\bar b$-dipole  with the target \cite{FZJL}.

\section{S-wave: $\ket{W}\to \ket{ t \bar b, L=0}$}
Once again for soft gluons  the azimuthal averaging leads to
\begin{equation}
\langle V_S(m_t,m_b)\rangle\simeq
{1\over Q^2}\left\{2Q^2z(1-z)
+(m_t-m_b)\left[(1-z)m_t-zm_b\right]\right\}^2
{2\bkappa^2\bk^{2}\over (\bk^{2} +\varepsilon^{2})^4}
\label{eq:VS}
\end{equation}
and
$
\langle A_S(m_t,m_b)\rangle=\left(g_A/ g_V\right)^2V_S(m_t,-m_b).$
 The 
S-wave term in (\ref{eq:FL})
 integrated over $\bk$ has 
approximately uniform $z$-distribution.
Then 
  the DLLA  estimate is as follows
\bea
F_L^S(x,Q^2)\simeq {2\alpha_S(\overline{\varepsilon^{2}})
\over 3\pi}
G(x,\overline{\varepsilon^{2}}),
\label{eq:1.11}
\eea
where  
$\overline{\varepsilon^{2}}\simeq (Q^2+2m_t^2)/4.$ 

\section{Numerical estimates}
To DLLA 
the CCNC contribution to $\sigma^{\nu N}$ with 
the gluon density $G(x,k^2)$ from \cite{IN2003}
is estimated as
$\sigma_{CCNC}^{\nu N}\simeq 0.43\times 10^{-31}$ cm$^2$
for 
$E_{\nu}=10^{12}\, $GeV. We neglected here the contribution
 of  hard gluons to the proton longitudinal  structure function. 
Therefore, the DLLA gives  the lower estimate for $F_L$.

For comparison, the frequently used 
massless approximation
gives at $E_{\nu}=10^{12}$ GeV the cross section $\sigma^{\nu N}$
that  for  different 
gluon densities varies in the range \cite{HJM}
$$0.2\times 10^{-31}\,{\rm cm^2}< \sigma^{\nu N}
< 1.5\times 10^{-31}\,{\rm cm^2}$$

\section{Scales and saturation}
At small-$x$ the unitarity/saturation effect enters the game
 \cite{Kancheli73, NZ75}. 
In massless approximation the
 unitarity correction to 
$\sigma^{\nu N}$
was found to be a  $50$ per cent effect \cite{KK2003}.
In particular, it was shown that the 
unitarity effect  turns  
$\sigma_{CC}^{\nu N}\simeq 1.\times 10^{-31}$ cm$^2$
at $E_{\nu}=10^{12}$ GeV into 
$\sigma_{CC}^{\nu N}\simeq 0.5\times 10^{-31}$ cm$^2$.
The strength of the 
unitarity/saturation effect depends on the hardness scale of the process,
 the first higher twist correction is estimated as \cite{GLR}
$$\sim {\alpha_S(Q^2)\over Q^2} {G(x,Q^2)\over \pi R^2}$$ 
The CCNC hardness scale, { $m_t^2$}, is much ``harder'' 
than  the hardness scale  for the light flavor currents. The latter 
 is $\lsim m^2_W$.
Thus we conclude that the unitarity  affects strongly the light quark 
contribution to $\sigma^{\nu N}$ but leaves the CCNC term intact.

\bigskip

{\bf{Acknowledgments.}} 
The authors are indebted to A. Papa for his kind invitation to present 
this talk at Diffraction 2010. Thanks are due  to 
 the Dipartimento di Fisica dell'Universit\`a
della Calabria and the Istituto Nazionale di Fisica
Nucleare - gruppo collegato di Cosenza for their warm
hospitality while a part of this work was done.
The work was supported in part by the Ministero Italiano
dell'Istruzione, dell'Universit\`a e della Ricerca,  by
 the RFBR grant 09-02-00732 and by the DFG grant 436 RUS 113/991/0-1.

\end{document}